\documentclass[aps,prd,floatfix]{revtex4}
\usepackage{epsf}

\newcommand{\bee}{\begin{equation}}
\newcommand{\ee}{\end{equation}}
\newcommand{\beea}{\begin{eqnarray}}
\newcommand{\eea}{\end{eqnarray}}

\def\slash{\!\!\!\!/\,}

\begin{document}
\title{Eigenvalue Decomposition of Meson Correlators}

\author{Thomas DeGrand}
\affiliation{
Department of Physics, University of Colorado,
        Boulder, CO 80309 USA}

\date{\today}

\begin{abstract}
Euclidean space hadronic correlators are computed in quenched QCD at small quark mass
using truncations of quark propagators which include or exclude
low eigenvalue eigenmodes of the Dirac operator.
High modes provide the dominant contribution to parity averaged correlators,
especially at short distances. Differences of correlators of opposite
parity receive most of their contributions from low modes
and are much smaller in size than parity averages at short distances.
 The pion 
propagator in any correlator to which it couples receives a large contribution from
low modes, while the tensor meson correlator
receives a  tiny contribution from low eigenmodes.
\end{abstract}
\maketitle

\section{Introduction}
Are all hadrons alike? This question was asked most famously\cite{Novikov:xj}
in the context of QCD sum rules, but it appears repeatedly in phenomenological
descriptions of hadronic structure. In this work I investigate a
particular kind of similarity: how eigenmodes of the 4-dimensional Dirac operator
contribute to Euclidean space
hadronic correlators, and so (indirectly) how these
eigenmodes contribute to hadronic spectroscopy and couplings in different channels.
My results might
have implications for the questions of 
whether excited states of mesons show features of chiral
symmetry restoration (namely, that the masses of various states,
and their couplings to currents, show degeneracies),
 what the origin of these features might be, and
how excited states differ from low lying meson states.

Lattice simulations encode ``snapshots'' of the QCD vacuum, and could presumably
address questions about the composition of hadronic states. 
This subject has been investigated before \cite{DeGrand:2000gq,DeGrand:2001tm}.
The idea is to compare the behavior of hadronic correlators built from quark propagators
which correspond to a truncated set of eigenmodes of the massless Dirac operator
with that of correlators computed ``exactly.''
Low eigenmodes of the Dirac operator
saturate the pseudoscalar and axial vector correlators
at large distance and contribute somewhat to the nucleon channel, less so
the vector and Delta channels.
They also make the dominant contribution to the eta-prime channel, as expected
from the Witten-Veneziano formula \cite{DeGrand:2002gm}. In this work I revisit
these studies, with an eye toward contrasting the Dirac eigenmode
composition of the long-distance part of hadronic correlators with the short-distance parts,
as well as the composition of correlators which are sums and differences of parity 
partners  ($\gamma_5$ and $1$, $\gamma_i$ and $\gamma_i\gamma_5$).
All these calculations reveal the following qualitative features of mesons in quenched QCD:
Low eigenmodes of the Dirac operator do not affect the part of correlators where high-lying states
appear. These low modes are the ones  which determine the quark
condensate via the  Banks-Casher  \cite{Banks:1979yr} relation, as well as eigenmodes
``at the QCD scale'' (a few hundred MeV). Other ways of describing the properties of low modes
are that they are strongly influenced by chirality-mixing interactions, and that they are
the main contributors to the strong interaction observed in the pseudoscalar current-current
correlator.

High modes contribute strongly to hadron correlators
 at short distance, where the correlators would be  dominated by excited states.
They contribute less to the long distance part of meson correlators. They do not make a major
contribution to parity asymmetric correlators.
 In this sense, valence quarks in high-lying states
decouple from the condensate.
This also means that even for light quarks, chiral symmetry
breaking and confinement are not directly related.
These observations are rather simple, and perhaps one does not ``need'' 
lattice simulations to see them. I suspect that no phenomenologist will find them surprising
(although I suspect that different phenomenologists will be unsurprised for different reasons.)
Instanton models plus lattice simulations have also been used to
compute the amplitude for chiral mixing of  $q \bar q$ 
correlators \cite{Faccioli:2003qz,Faccioli:2003hb}.
Ref. \cite{DeGrand:2000gq} also observed that the lowest eigenvalue modes had
a large autocorrelation in both density and chirality at short distance,
which slowly dies away as
the eigenmode rises.

Analyses of hadronic spectroscopy 
\cite{Cohen:2001gb,Glozman:2002kq,Glozman:2002jf,Cohen:2002st,Glozman:2003mm,Beane:2002td,Glozman:2003gk}
have been used to argue that excited states of the meson and baryon spectrum are
effectively chirally restored, in the sense that particles of opposite parity are
nearly degenerate in mass.  It is argued that this ``effective restoration of chiral symmetry''
is a natural consequence of the soft nature of chiral symmetry breaking: generally
in quantum mechanics, low excitations are sensitive to symmetry breaking but high
excitations are not. For mesons, the crossover to chiral symmetry is
argued to occur in the range 1.5-2 GeV.

In perturbation theory, correlators of currents which are chiral partners of
each other
(the vector and axial vector currents, for example) are identical, because
QCD is a vector theory and chirality is conserved at all quark-gluon vertices.
Condensates allow one to parameterize nonperturbative physics, and can lead
to differences in correlators. However, QCD is a confining theory, and the
spectrum of QCD is one of bound states, so identity of the correlators
must have consequences for spectroscopy and matrix elements. That
is especially true for the large-$N_c$ limit of QCD, where all excitations
are narrow resonances. Arguments based on semi-local duality favoring effective
chiral restoration at high excitation have been made
\cite{Beane:2001uj}. but have been
 criticised as being cutoff dependent\cite{Golterman:2002mi}.

A summary of ideas supporting
chiral restoration and its connection to 
 the quark model has recently been given by Swanson\cite{Swanson:2003ec}.

In instanton models\cite{Schafer:1996wv} low eigenmodes of the Dirac operator are built from
 an overlap
of single-instanton zero modes, and are strongly chirally asymmetric. As the magnitude
of the quark eigenmode rises, the eigenfunctions couple less and less to instantons, and 
there should be a
``crossover'' to chirally-symmetric physics.

The common lattice correlators used to extract spectroscopy involve projections onto
 momentum eigenstates.
For example, the $\vec p=0$ correlator
is found by by averaging over time slices:
\beea
C_j(t)&=&
 \sum_x  \langle 0 | O_j(x,t)O_j(0,0) |0\rangle 
\nonumber \\
  &=& \sum_n {{ \langle 0 | O_j | n \rangle|^2}\over{2m_n}} \exp(-m_n t)
\nonumber  \label{eq:ct}  \\
\eea
A second kind of observable is just the point-to-point correlator
\bee
\Pi_i(x) = {\rm Tr}\langle J_i^a(x)J_i^a(0)\rangle,
\label{eq:pt}\ee
where the current will be proportional to
$J_i^a(x) = \bar \psi(x) {\tau^a } \Gamma(i) \psi(x)$, and its ratio
to the free-field correlator
$R_i(x) = \Pi_i(x)/\Pi_i^0(x)$.
It is very difficult to compute masses of highly excited states from lattice simulations.
Their signals vanish exponentially compared to (and underneath) the lightest state in the channel.
Each new state requires a fit with two more parameters (the mass and $\langle 0 | O_j | n \rangle|^2$),
which in turn requires a fine lattice with many lattice points. At very short distance (order
one lattice spacing) lattice discretization artifacts affect results. (Note that correlators like
$C(t)$ probe long distances, even at small $t$.)
I have not done direct calculations of spectra,  and so my approach is more indirect and my
results are tentative.

 I will look at correlators with point sources and sinks,
 $O_j(x,t) = \bar \psi(x,t) \Gamma_j \psi(x,t)$.
Lattice simulations usually do not use pointlike currents as interpolating fields
because they do not couple to low states as well as more extended operators,
 and parameters of the lowest
states are usually the goals of the simulation.
However, for 
our purposes, it will be interesting to keep this simple form, and look at
$\Gamma_j$'s which are chiral partners.

In contrast to usual lattice simulations, all results presented here are qualitative.

\section{Exact Results from Overlap Actions}
The calculations presented here are done with overlap \cite{ref:neuberfer}
fermions. This lattice
discretization preserves a lattice version of exact chiral symmetry
at nonzero lattice spacing, without flavor doubling, making it particularly useful for
addressing questions associated with chiral symmetry. The following properties
of overlap actions are relevant to this work:

 The eigenmodes of any massless
overlap operator are located on a circle in the complex plane of radius $x_0$
with a center at the point $(x_0,0)$. The corresponding eigenfunctions are
either chiral (for the eigenmodes with real eigenvalues
located  at $\lambda=0$ or $\lambda=2x_0$)
or nonchiral  and paired; the two eigenvalues of the paired nonchiral modes
are complex conjugates.

 The massive overlap Dirac operator for bare quark mass $m$ is
conventionally defined to be
\bee
D(m) = ({1-{m \over{2x_0}}})D(0) + m
\ee
and it is also conventional to define the propagator so that the chiral
modes at $\lambda=2x_0$ are projected out,
\bee
\hat D^{-1}(m) = {1 \over {1-m/(2x_0)}}(D^{-1}(m) - {1\over {2x_0}}) .
\ee
The contribution to the propagator of a single (positive chirality)
zero mode in the basis
where $\gamma_5= {\rm diag}(1,-1)$ is
\bee
\hat D(m)^{-1}= {1\over m}\pmatrix{ |j+\rangle\langle j+|   &0\cr 0 & 0 \cr}.
\ee
Nonzero eigenvalue eigenmodes of $D(0)^\dagger D(0)$ are also chirality eigenmodes.
The $j$th pair of nonchiral modes contributes a term to the propagator
\bee
\hat D(m)_j^{-1}= \pmatrix{ \alpha_j |j+\rangle\langle j+| & -\beta_j|j+\rangle\langle j-|\cr
 \beta_j |j-\rangle\langle j+|& \alpha_j |j-\rangle\langle j-|\cr},
\label{eq:dmmode}
\ee
where, using
$D(0)^\dagger D(0) |jh\rangle = \lambda^2_j |jh\rangle$ for chirality $h$, 
$\mu=m/(2x_0)$, and $\epsilon_j= \lambda_j/(2x_0)$, the entries are
\bee
\alpha_j={1\over{2x_0}}
{{\mu(1-\epsilon_j^2) }\over{\epsilon_j^2 + \mu^2(1-\epsilon_j^2)}}
\ee
\bee
\beta_j={1\over{2x_0}}
{{\epsilon_j\sqrt{1-\epsilon_j^2})  }\over
{\epsilon_j^2 + \mu^2(1-\epsilon_j^2)}} .
\ee
(The eigenmodes of $D(0)$
have eigenvalues $2x_0(\epsilon_j^2 \pm i\epsilon_j\sqrt{1-\epsilon_j^2})$.)
For a summary of these (and other) useful formulas, see Ref. \cite{ref:FSU98}
(for the special case $x_0=1/2$).

What is important about these well-known results is their application to
correlators expressing the sum and difference
of opposite parity channels:
\bee
C_{\pm\Gamma}(x,x') \propto {\rm Tr} \hat D^{-1}(m) \Gamma  \Delta_\pm  \Gamma
\ee
where
\bee
\Delta_\pm =   \hat D^{-1}(m) \mp \gamma_5 \hat D^{-1}(m) \gamma_5
\ee
The parity difference is particularly simple: it involves the sum combination,
which  because of the Ginsparg-Wilson relation \cite{ref:GW} is equal to
\bee
\Delta_- =   
 2m\hat (D^{-1}(m))^\dagger \hat D^{-1}(m)   
\ee
The volume-summed trace of $\Delta_-$ gives the Gell-Mann-Oakes-Renner (GMOR) relation.
 Eigenmode by eigenmode, $\Delta_+$ and
$\Delta_-$ can be evaluated directly from Eq. \ref{eq:dmmode}; the former
 contribution comes from
the off-diagonal part of $\hat D^{-1}$ while the latter contribution comes from the
diagonal elements.

This decomposition makes no reference to any dynamics. Thus
the parameter which differentiates between chiral symmetry and asymmetry is just
the quark mass--it is the only chiral breaking parameter which is available.
The ratio of a contribution of an eigenmode to the chiral-difference correlator
to the chiral symmetric correlator scales roughly as $m/\lambda_j$. This is rather similar
to continuum free-field behavior, with $\Delta_-(p) = m/p^2$, $\Delta_+(p) = p\slash/p^2$.
It is NOT the behavior seen for Wilson-like fermions, which have an explicit
chiral-symmetry breaking term at large momentum,
 $D^{-1}(m) = m + i \sum_\mu\gamma_\mu \sin(p_\mu a)/a + 2 a \sum_\mu \sin^2(p_\mu a/2)$.
Staggered fermion correlators entangle parity partners in a way I do not know how
to present simply.

Naively, this means that, unless the $m\rightarrow 0$ limit is singular,
parity difference correlators will go to zero with the quark mass.
However, we have to be more careful \cite{Leutwyler:1992yt}, because the 
eigenmode density $\rho(\lambda)$ diverges in the ultraviolet as $\lambda^3$.
Mode sums  generally require subtraction:
high eigenmodes do contribute to all susceptibilities.
For example, the pseudoscalar susceptibility  (time integral of $C_j(t)$ for
$\Gamma_j=\gamma_5$) is
\bee
\chi_\pi = {1\over V} \sum_{x,y} \langle \pi^a(x)\pi^a(y) \rangle = 
 \int d\lambda \rho(\lambda){1 \over {\lambda^2 + m^2}}
\ee
and the scalar susceptibility is
\bee
\chi_{a_0} = {1\over V} \sum_{x,y} \langle a_0^a(x)a_0^a(y) \rangle = 
 \int d\lambda \rho(\lambda){{\lambda^2 - m^2} \over {(\lambda^2 + m^2)^2}}
\ee
For either susceptibility, it is necessary
to break the integral over $\lambda$  into two parts, one for the low modes
$0<\lambda<\Lambda_c$ where $\Lambda_c >>m$, and one for the high modes. The low
modes give the Banks-Casher relation for $\chi_\pi$, $\chi_\pi=\rho(0)/(\pi m)+\dots$. The
high frequency part of the integral must be subtracted twice,
\bee
\chi_\pi =
 {1\over m} \big( \int_0^{\Lambda_c} d\lambda \rho(\lambda){m \over{\lambda^2+m^2}} \big)
+ \gamma_0  + \gamma_2 m^2  + m^4 K_\pi(\Lambda_c,m).
\label{eq:cps}
\ee
The scalar susceptibility has a similar behavior, except that there is no contribution from
$\lambda=0$ \cite{Smilga:1993in}.
For overlap fermions, the GMOR relation is exact,
\bee
\chi_\pi = {2\over m} \langle \bar \psi \psi(m) \rangle
\ee
and the scalar susceptibility is
\bee
\chi_{a_0} = 2 {d \over {dm}}\langle \bar \psi \psi (m)\rangle ,
\ee
so that the expansion coefficients in the expansions for $\chi_\pi$ and $\chi_{a_0}$ are related.
In the difference $\chi_\pi - \chi_{a_0}$ the contribution from the chiral condensate
(which comes from eigenmodes near $\lambda=0$)
survives and dominates in the $m\rightarrow 0$ limit.
The mass-independent term in $\chi_\pi - \chi_{a_0}$ cancels.  The chiral sum 
$\chi_\pi + \chi_{a_0}$ retains the high-$\lambda$ dependent $\gamma_0$ coefficient.
The terms proportional to positive powers of $m$  vanish in the massless limit in either case.
What is true for an integral is not necessarily true for an integrand,
but it is a plausible assumption to expect that chiral difference
correlators themselves will receive large contributions only from low eigenmodes,
while high eigenmodes will contribute strongly to chiral sum correlators.
For example, one might expect that the low modes would dominate the pseudoscalar correlator:
then \cite{Golterman:2003qe},
\bee
C_{PS}(x,x')= \langle \{ \sum_\lambda \langle j,\lambda, x|j,\lambda, x\rangle
{1\over{\lambda^2+m^2}}\langle j,\lambda, x'|j,\lambda, x'\rangle \}\rangle
\ee
This reasoning is dangerous because of the UV divergence of $\rho(\lambda)$, but it
can be tested by simulation.

Zero modes decouple from $C_+(x,x')$.  They make a large contribution to $C_-(x,x')$.
This is not surprising: zero modes are chiral. They will, however, considerably
distort chiral-difference correlators in small volumes.

\section{Examples}

\begin{figure}[!thb]
\begin{center}
\epsfxsize=0.6 \hsize
\epsffile{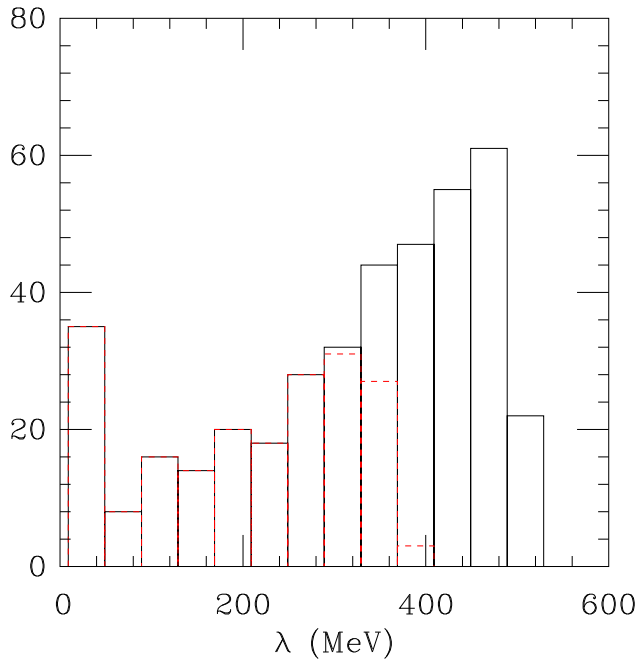}
\end{center}
\caption{
Histogram of eigenvalues of the Dirac operator of eigenmodes recorded for this study.
The peak at the lowest box contains all zero modes. The 
black 
solid lines
curve shows the distribution of
the twenty lowest eigenmodes of $D^\dagger D$, while the 
red 
dashed lines separate the distribution of
ten modes.
 }
\label{fig:hist}
\end{figure}

I have done  a set of quenched spectroscopy runs using a particular implementation
of the overlap operator \cite{ref:TOM_OVER}  with HYP \cite{ref:HYP} gauge connections.
Eigenmodes of the Dirac operator are computed using the conjugate gradient algorithm
of Ref. \cite{ref:eigen}. 
These studies are done on a small data set of quenched configurations. 
It has 20 $16^4$ lattices generated with the Wilson gauge action at coupling
$\beta=6.1$, corresponding to  a lattice spacing of about 0.09 fm.  These lattices have
the same lattice spacing and quark masses as were used in a large scale matrix element
simulation   \cite{DeGrand:2003in},
so all necessary hadron masses and matrix elements are in principle known. I have
computed the lowest 20 eigenmodes of $D(0)^\dagger D(0)$ and recoupled them into
eigenmodes of $D$: there are $N_0$ chiral zero modes and $2(20-N_0)$ nonchiral
paired modes. Eigenmodes of the overlap Dirac operator whose eigenvalue is less than 
around 500  MeV have been computed. The spectrum of the imaginary part of the
 eigenvalue
is shown in Fig. \ref{fig:hist}. The energy is inferred from the lattice data assuming
an inverse lattice spacing of 2200 MeV.
I compute hadron correlators using full quark propagators,
and several kinds of quark propagators built of truncated mode sums:
quark propagators with only zero modes, propagators without zero modes, and
propagators from which the lowest 10 or 20 eigenmodes of $D^\dagger D$ have been excluded.

I will show results corresponding
to bare quark masses in lattice units of $am_q=0.020$, which is about half the
strange quark mass. A  meson made of a pair of $am_q=0.020$ quarks is an approximation
to the physical kaon. Results for other small quark masses are similar.
 At higher quark masses (above $am_q=0.050$ or $m_q \ge 2.5 m_s$) the qualitative features I
present do no longer occur. Pictures shown
 in Refs. \cite{DeGrand:2000gq,DeGrand:2001tm} show that low
modes cease to contribute a dominant part of  any correlator.

 Because these
simulations are done in finite volume, and in quenched approximation,
some channels have contributions from zero modes of the Dirac operator. These
contributions would not be present in infinite volume. In other channels these modes
are absent.

\subsection{``Spectroscopic'' correlators with point sources}

\begin{figure}[!thb]
\begin{center}
\epsfxsize=0.7 \hsize
\epsffile{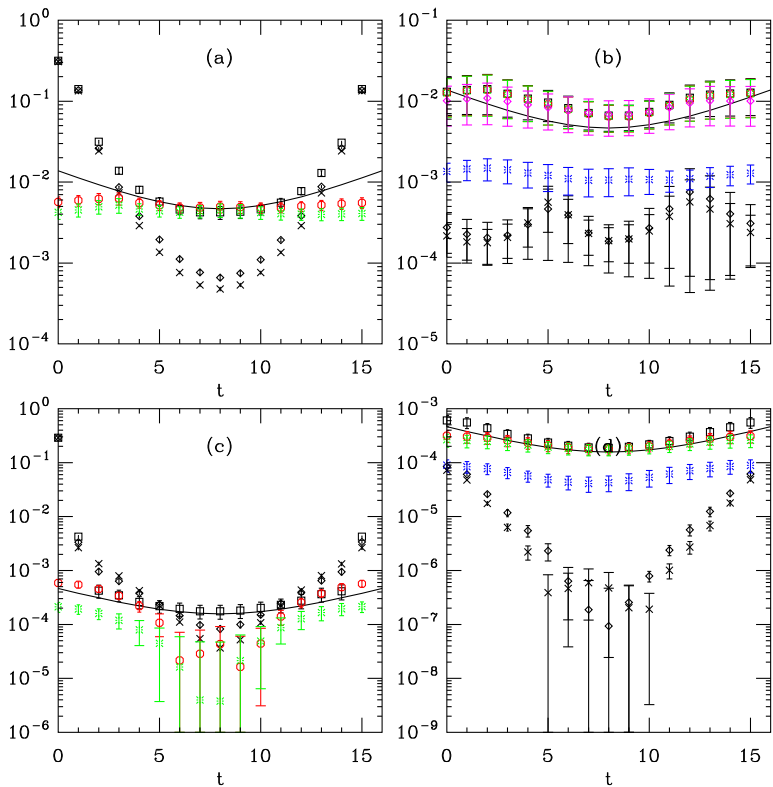}
\end{center}
\caption{
Combinations of point-to-point
pseudoscalar and scalar ($\pi \pm a_0$) sum (a) and difference (b)
correlators,
and axial vector $\gamma_0\gamma_5-\gamma_0\gamma_5$ and scalar $\gamma_0-\gamma_0$
sums (c) and differences(d). The full correlators are shown by open squares.
Thheir approximation by restricted-mode
quark propagators with
20 lowest modes 
(red octagons)
 or 10 lowest modes 
(green bursts).
Crosses and diamonds show the contribution to the correlator
from quark propagators with 20 and 10 low modes excluded.
The curve is the contribution to the correlator
from the pion, using its measured mass and decay constant from
a different simulation.
In panels (b) and (d) the  contribution of the lowest 20 eigenmodes
 with zero modes excluded is shown in
blue fancy diamonds, 
and the  pure zero mode contribution 
(magenta fancy crosses) is
shown in (b).
 }
\label{fig:comppions}
\end{figure}

I begin first with correlators $C(t)$ of Eq. \ref{eq:ct}.
Results for pseudoscalars and scalars are shown in Fig. \ref{fig:comppions}. Panel (a)
shows the sum of pseudoscalar and scalar correlators.
Note the complete saturation of the correlator for $t>5$ by low modes,
and the essentially complete saturation of
the correlator by high modes at shorter distances. The  pionic contribution
to the correlator, using $\langle 0 |  \bar \psi \gamma_5 \psi | PS\rangle = m_{PS}^2 f_{PS}/(2m)$,
(with the fitted parameters $m_{PS}$ and $f_{PS}$ 
 from Ref. \cite{DeGrand:2003in})  is superimposed.
The low modes make up the bulk of the pion's contribution to the correlator.

Panel (b) shows the difference of pseudoscalar and scalar correlators.
This signal is heavily influenced by zero modes, so that while low modes again contribute
a ``pion-like'' signal, most of this contribution involves at least one zero mode.
High eigenmodes make a negligible contribution to the correlator at any time
separation.

The axial-scalar $(\gamma_0\gamma_5-\gamma_5\gamma_0) \pm (\gamma_0-\gamma_0)$
combination shows similar behavior (panels (c) and (d).
The axial vector current decouples from the Goldstone mode in the chiral
limit, and this is reflected in the ``sum'' correlator by
a slower saturation of the pion contribution to the correlator by low
modes.
Again, the bulk of the signal at small time steps comes from high modes.
High modes do not contribute to the chiral difference.

\begin{figure}[!thb]
\begin{center}
\epsfxsize=0.7 \hsize
\epsffile{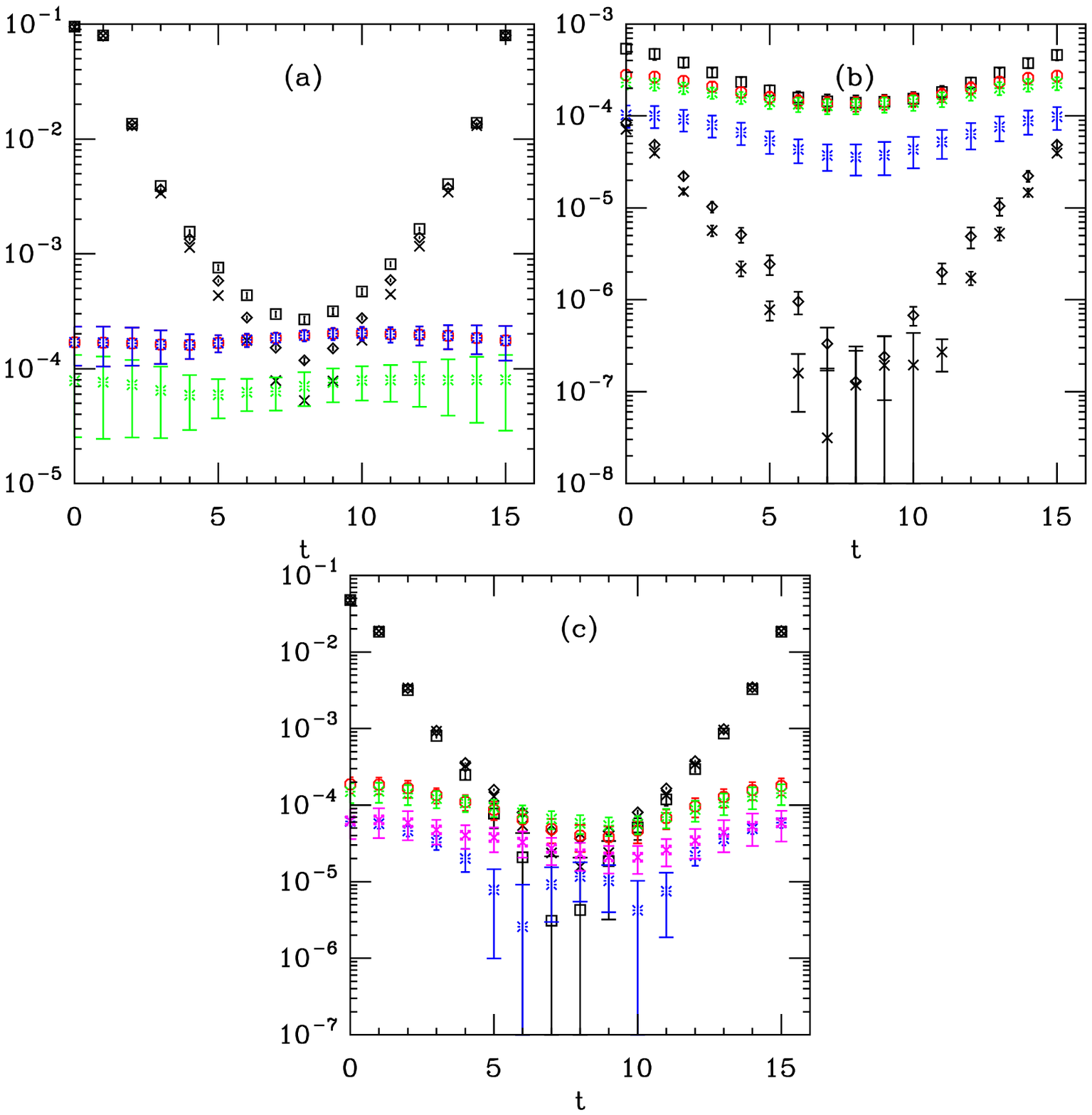}
\end{center}
\caption{
Point-to-point
$(\gamma_i-\gamma_i) \pm (\gamma_i\gamma_5-\gamma_i\gamma_5)$ (rho plus/minus $a_1$)
correlator: (a) sum, (b) difference,  and (c) tensor or $a_2$
correlator
$(\gamma_i\gamma_j - \gamma_i\gamma_j)$.
Full correlators are shown as black squares, and their approximation by restricted-mode
quark propagators with
20 lowest modes 
(red octagons),
10 lowest modes 
(green bursts), and pure zero mode contribution 
(magenta fancy crosses)
are also shown.
The  contribution of the lowest 20 eigenmodes
 with zero modes excluded are  
blue fancy diamonds.
Crosses and diamonds show the contribution to the correlator
from quark propagators with 20 and 10 low modes excluded.
 }
\label{fig:compvat}
\end{figure}

Vector and axial combinations (Fig. \ref{fig:compvat}, panels (a) and (b))
lack any single state which is dominated by low modes. High modes
again make a tiny contribution to the chiral-asymmetric channel, and saturate the signal
in the chirally symmetric channel. The absolute size of the correlators in the
two channels is quite different at short distances.

The correlator $C_+(t)$ is almost two orders of magnitude larger than 
$C_-(t)$ in the vector-axial vector channel, and almost three orders of magnitude
larger in the pseudoscalar-scalar channel. This is not a direct observation that the
spectrum of quenched QCD is parity-doubled at high excitation.
However, this   is the ``raw data'' from which
masses are extracted, so it is likely that any fit to it for masses
will produce parity symmetric results.

\begin{figure}[thb]
\begin{center}
\epsfxsize=1.0 \hsize
\epsffile{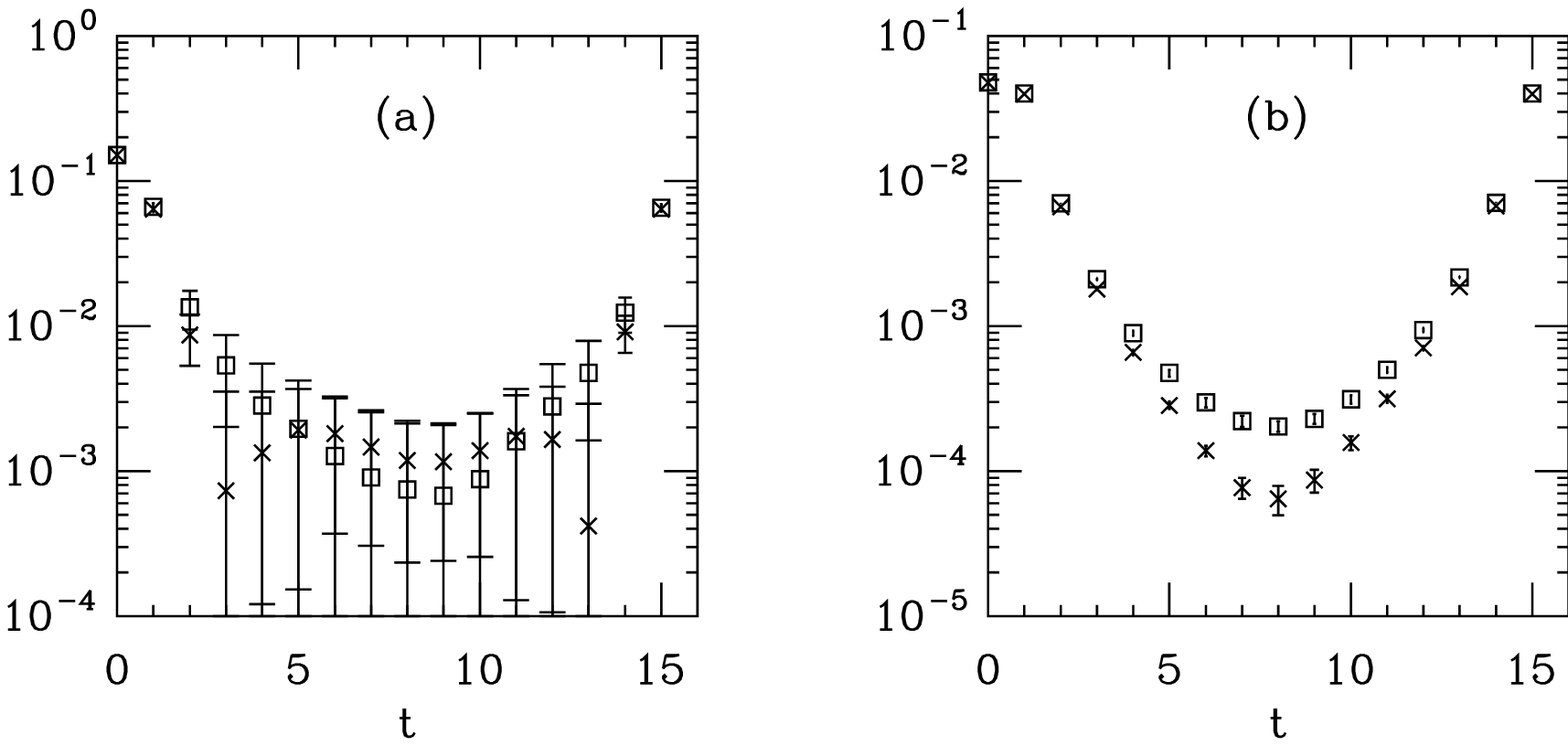}
\end{center}
\caption{ (a)
Point-to-point
$(\gamma_5 - \gamma_5)$ correlator, with pionic contribution removed
(squares) and $1-1$ (crosses) correlators at bare mass $am_q=0.02$.
(b) point-to-point
$(\gamma_i - \gamma_i)$ (squares) and $(\gamma_i\gamma_5 - \gamma_i\gamma_5)$ (crosses) correlators
 }
\label{fig:compdiff}
\end{figure}

Perhaps direct comparisons, in Fig \ref{fig:compdiff}, are clearer.
Panel (a) shows the pseudoscalar and scalar correlators; I have subtracted the measured
pion contribution from the pseudoscalar channel. Panel (b) shows the
vector and axial vector correlators.
The correlators $C_j(t)$ defined in Eq. \ref{eq:ct},
where $\Gamma_j$ are parity partners ($\gamma_5$ and $1$, $\gamma_i$ and $\gamma_i\gamma_5$),
become equal for $t$ less than about 0.2 fm.

\subsection{Point-to-point correlators}
\begin{figure}[!thb]
\begin{center}
\epsfxsize=0.7 \hsize
\epsffile{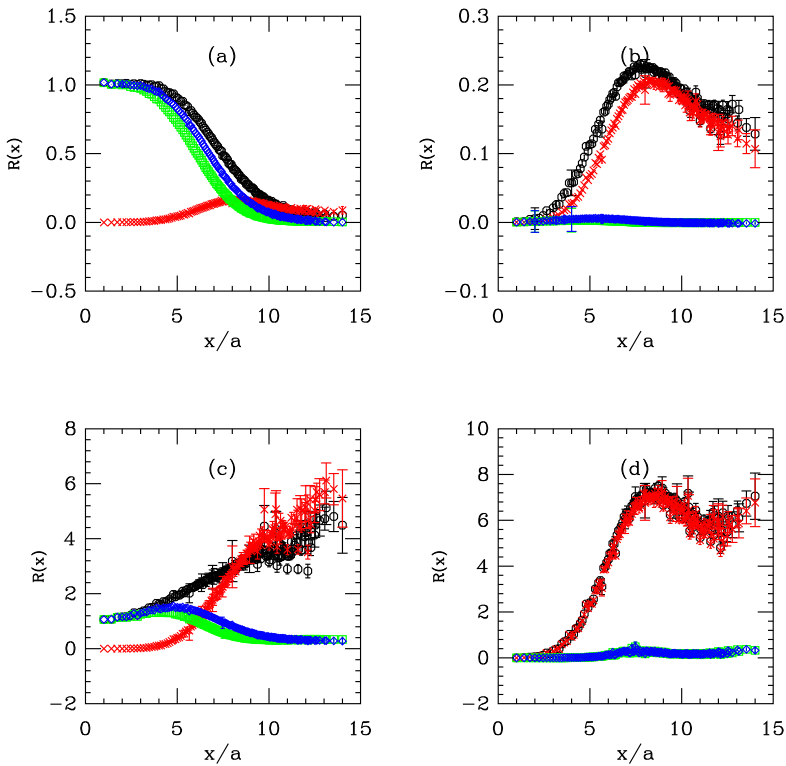}
\end{center}
\caption{
Point-to-point
$(\gamma_\mu-  \gamma_\mu)\pm (\gamma_\mu\gamma_5-  \gamma_\mu\gamma_5)$
vector-axial vector sum (a) and difference (b), and
$(\gamma_5-  \gamma_5)\pm (1-1)$ pseudoscalar-scalar sum (c) and difference (d), 
normalized by  free field vector and scalar currents, Full correlators are shown in black
 and  approximation by restricted-mode
quark propagators with
20 lowest modes 
(red crosses)
and contributions from propagators containing all but the lowest 20 
(blue diamonds) or 10 
(green squares) modes.
 }
\label{fig:ptp}
\end{figure}

One can also compute point-to-point correlators (Eq. \ref{eq:pt}).
 The  short distance, parity-summed correlators
receive little contribution from low eigenmodes.
That their ratios approach unity at small x is just asymptotic freedom.
However, this is a confining theory: the spectrum of quenched QCD consists of
zero width resonances plus quenched artifacts (mostly associated with eta prime
hairpins).
A more correct statement of what is seen is that the
 resonances which make up the low-x part of the sum correlators
receive little contribution
from low modes. Results are shown in Fig. \ref{fig:ptp}.

The parity difference correlators vanish at low x.  Low eigenmodes (including zero modes) 
dominate the correlators at large x, where the lightest states in the channel appear.
Alternatively, the strong attraction seen at large $x$ in the pseudoscalar channel
comes almost entirely from eigenmodes below 500 MeV.

\begin{figure}[!thb]
\begin{center}
\epsfxsize=0.6 \hsize
\epsffile{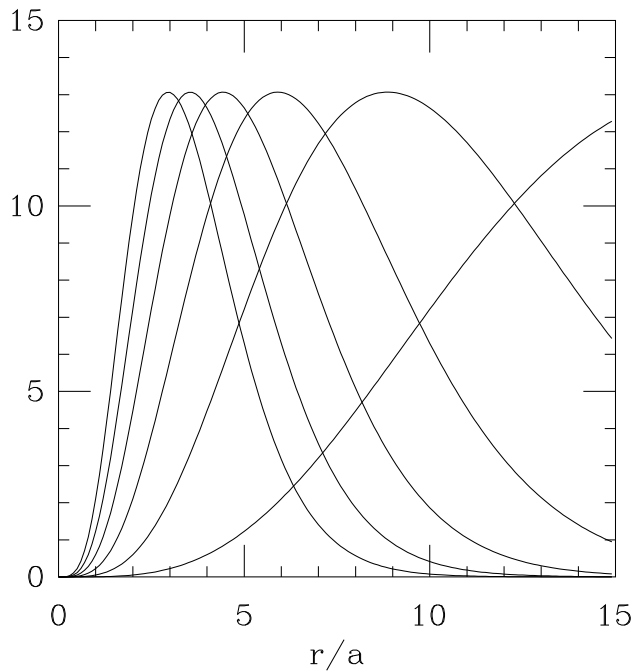}
\end{center}
\caption{
$x^5 K(x)$, the contribution of a resonance of mass $m$ to one of the
 point-to-point correlator ratios.
The peaks correspond to lattice masses (from left to right) of $m_Ha=1.5$, 1.25, 1.0, 0.75, 0.5,
and 0.25 (the peak of the last curve lies outside the graph).
 }
\label{fig:ratiolat}
\end{figure}

In the absence of reliable spectroscopy calculations, we can roughly
quantify the ``breakpoint''
between low and high mode contributions by considering the contribution of a single
resonance of mass $m$ to one of these ratios. This curve is proportional to
the ratio $x^5 K(m,x)$, where
$K(m,x)$ is the free field propagator for a particle of mass $m$. A family of
curves of varying masses (in lattice units) is shown in Fig. \ref{fig:ratiolat}; the
height has been rescaled as $m^2$ to produce a plateau of maxima. What is important
from this picture is not the height of the peaks, it is their location. If we compare
the high eigenmode and full propagator curves in Fig \ref{fig:ptp}(a) and (c),
we see that they separate at a lattice distance of about 5 units. Any resonance with
a lattice mass lighter than $am_H< 0.75$ would make  its peak contribution at larger $x$.
This mass is about 1700 MeV, and is the rough dividing point between hadrons built dominantly of
chiral sensitive modes and ones which are not, for quenched QCD.
(Again, recall that the  inverse lattice spacing
of these simulations, above which discretization effects dominate physics, is about 2.2 GeV).
Swanson's estimate \cite{Swanson:2003ec} of 2.5 GeV for the crossover corresponds
 to $ma \simeq 1.1$,
but given the roughness of either estimate, I would not take the difference seriously.

\begin{figure}[!thb]
\begin{center}
\epsfxsize=0.6 \hsize
\epsffile{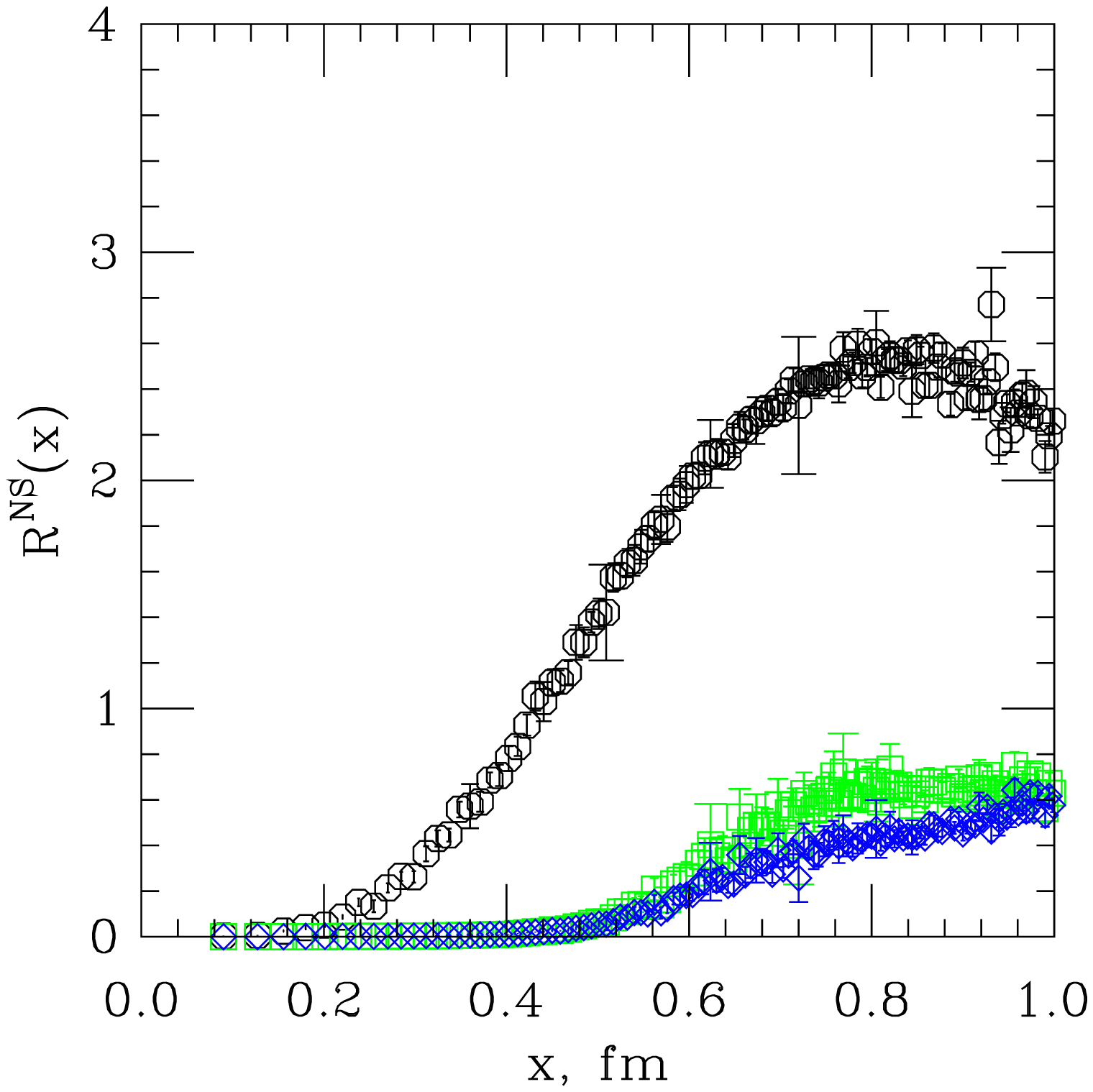}
\end{center}
\caption{
$R^{NS}(x)$ as defined in Eq. \protect\ref{eq:flip} and its contribution from propagators
from which the lowest 20
 (blue diamonds) and 10 (green squares)
 eigenmodes of $D^\dagger(0)D(0)$ are excluded.
 }
\label{fig:nf}
\end{figure}

Finally, we show in Fig. \ref{fig:nf}
the saturation of the amplitude for chirality flipping proposed by
Ref. \cite{Faccioli:2002xf} and computed on the lattice in Ref. \cite{Faccioli:2003qz},
\bee
R^{NS}(x) = {{\Pi_\pi(x)-\Pi_{a_0}(x)}\over {\Pi_\pi(x)+\Pi_{a_0}(x)}},
\label{eq:flip}
\ee
using complete quark propagators and high-eigenmode truncations. Clearly
the high eigenmodes make a tiny contribution to $R^{NS}$;
it is the low modes that are strongly influenced by chirality mixing interactions.

\section{Conclusions}

Lattice simulations in which contributions to quark propagators from low and high
eigenvalue eigenmodes of the Dirac operator are separated identify
 the following qualitative features of light-quark mesons in quenched QCD:
Low eigenmodes make a large contribution to the pion propagator. They are
responsible for the strong long-distance attractive
 interaction seen in the pseudoscalar channel.
Low eigenmodes of the Dirac operator make a small contribution to the short distance part
of correlators (where excited states contribute).
These are the eigenmodes which determine the quark
condensate via the  Banks-Casher relation, as well as eigenmodes
at the QCD scale'' (a few hundred MeV). This implies that valence quarks in high-lying states
decouple from the condensate.

I have not directly observed parity doubling
 at high excitation in the meson spectrum, but lattice
 correlation functions in parity-partner channels
(from which fits to masses would be performed) are equal at a part in $10^{3-4}$
at short distances. Point-to-point correlators suggest that hadrons of mass
above 1.7 GeV are insensitive to low Dirac eigenmodes. 
 Low modes do not contribute to the tensor meson correlator.

Perhaps these results can be used to constrain phenomenological models of
hadron structure.
Results such as Fig. \ref{fig:comppions}(a) suggest that it might be profitable for lattice
calculations done at small quark mass
to exploit low eigenmodes of the Dirac operator in simulations.

\section*{Acknowledgments}
This work arose through discussions with Leonid Glotzman, and I am grateful to him
for numerous conversations, correspondence, and encouragement.
This project was
begun while I was a guest at the Max Planck Institute for Physics and
Astrophysics, Munich, and I appreciate for that institution's hospitality.
This work was supported by the US Department of Energy.

\end{document}